\documentclass[11pt]{article}
\usepackage[utf8]{inputenc}
\usepackage[letterpaper,margin=1in]{geometry}
\usepackage[backend=biber,citestyle=phys,bibstyle=physplus,maxnames=3]{biblatex}
\usepackage{physics}
\usepackage{siunitx}
\usepackage{hepnicenames}
\usepackage{subcaption}
\usepackage{todonotes}
\usepackage{wrapfig}

\usepackage{tikzexternal}
\tikzsetexternalprefix{generated-figures/}
\colorlet{partondist}{green!60!black}
\colorlet{gluondist}{red!80!black}
\colorlet{fragmentation}{cyan!60!black}
\colorlet{kinfactor}{blue!40!red!70!black}

\newcommand\h[4]{\mathcal{H}^{(#1)}_{#2#3#4}}

\newcommand\F[5]{F^{(#2)}_{#1#3#4#5}}

\newcommand\pA{\ensuremath{\mathrm{p}A}}

\newcommand\xprojectile{\ensuremath{x_p}}
\newcommand\xtarget{\ensuremath{x_g}}
\newcommand\zhadron{\ensuremath{z}}

\newcommand\kperp{\ensuremath{k_\perp}}
\newcommand\pperp{\ensuremath{p_\perp}}

\newcommand\rapidity{\ensuremath{y}}

\newcommand\dipoleF{\ensuremath{\mathcal{F}_{\xtarget}}}

\newcommand\etal{\textit{et al}}

\DeclareRobustCommand{\Pnucleus}{\HepGenParticle{A}{}{}\xspace}
\DeclareRobustCommand{\Panything}{\HepGenParticle{X}{}{}\xspace}
\DeclareRobustCommand{\Phadron}{\HepGenParticle{h}{}{}\xspace}

\title{Saturation Physics on the Energy Frontier}
\author{David Zaslavsky}
\begin{document}
 \begin{titlepage}
  \begin{flushright}
   DPF2015-255\\
   \today
  \end{flushright}
  {
   \let\newpage\relax%
   \maketitle%
  }

  \begin{abstract}
   Saturation physics is expected to be relevant at sufficiently small parton momentum fractions $x$ in high-energy proton- (or deuteron-)ion collisions at RHIC and the LHC. Accordingly, these collisions provide the best available testing ground for the saturation model. However, producing precise numerical predictions from the model is a complicated task; the state of the art in this area involves next-to-leading order QCD calculations, which are difficult to do numerically. Here I'll review recent progress in extracting numerical predictions from saturation models and matching them to experimental results.
  \end{abstract}
  \begin{center}
   \textsc{Presented at}
   
   \vspace{2em}\Large
   DPF 2015\\
   The Meeting of the American Physical Society\\
   Division of Particles and Fields\\
   Ann Arbor, Michigan, August 4--8, 2015
  \end{center}
 \end{titlepage}
 
 \section{Saturation and Hybrid Factorization}
 Almost since the dawn of QCD, the behavior of gluons at small longitudinal momentum fractions $x$ has been an area of interest among hadron structure researchers. The seminal work in the field is of course the BFKL evolution equation~\cite{Fadin:1975cb,Balitsky:1978ic}, which predicts that gluon distributions should rise sharply with momentum fraction as $x$ becomes smaller. However, it was quickly realized that this growth is unsustainable, and in fact violates the Froissart bound (or equivalently, violates unitarity) at sufficiently small $x$. Adding a nonlinear term to the equation fixes this problem by tempering the growth of the gluon distribution~\cite{Gribov:1981ac,Balitsky:2001re,Kovchegov:1999ua}, producing an asymptotic approach to a constant gluon field strength as $x\to 0$. This is the phenomenon of gluon saturation. Physically, saturation represents gluon recombination within the target wavefunction, or multiple gluon scattering by the projectile parton, depending on the time-ordering of the relevant processes.
 
 \begin{wrapfigure}{R}{6cm}
  \tikzsetnextfilename{pAcollision}
  \begin{tikzpicture}[scale=0.7,every node/.append style={scale=0.75}]
   \node[blob] (junction1) at (-2,2) {};
   \node[blob] (junction2) at (-2,-2) {};
   \node[interaction] (junction3) at (0,0) {};
   \node[blob] (junction4) at (2,0) {$\zhadron$};
   \draw[baryon] (junction1) +(-1,0) to (junction1);
   \path (junction1) +(-1,0) node[left] {$\Pproton$};
   \draw[baryon] (junction1) to +(2,1);
   \draw[quark] (junction1) ++(0,-\baryonlinespacing) coordinate (junction1R) -- (junction3) node[above right=1pt,pos=0.5,momentum] {$\xprojectile p_p$} -- (junction4) node[pos=0.5,below,momentum] {$\kperp$};
   \draw[quark] (junction4) -- +(30:2) node[pos=0.7,above left=1pt,momentum] {$\pperp$} node[pos=1,above right] {$h$};
   \draw (junction4) to +(-25:2);
   \draw (junction4) to +(-35:2);
   \draw (junction4) to +(-30:2) node[below right] {$X$};
   \draw[nucleus] (junction2) +(-2,0) to (junction2) to +(2,-1);
   \path (junction2) +(-2,0) node[left] {$A$};
   \draw[gluon] (junction2) -- (junction3) node[below right,pos=0.5,momentum] {$\xtarget p_A$};
   \node[fill=partondist!60!white,blob] at (junction1) {};
   \node[fill=gluondist!60!white,blob] at (junction2) {};
   \node[draw,fill=kinfactor!60!white,interaction] at (junction3) {};
   \node[fill=fragmentation!60!white,blob] at (junction4) {$\zhadron$};
  \end{tikzpicture}
  \caption{Schematic of the hybrid factorization}
  \label{fig:pAcollision}
 \end{wrapfigure}
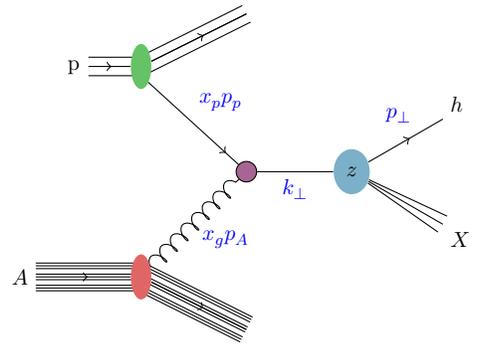
 
 However, the physical effect of saturation has yet to be detected experimentally.
 In other words, currently available experimental data can't distinguish between models that include saturation and those that do not, and also can't constrain the parameters of the saturation model.
 Proton- (or deuteron-)nucleus collision cross sections from the LHC and RHIC may change that, though.
 The $\HepProcess{\Pproton\Pnucleus\to\Phadron + \Panything}$ cross section in the forward region, as shown in figure~\ref{fig:pAcollision}, is particularly sensitive to the behavior of small-$x$ gluons in the nuclear target.
 With a sufficiently precise calculation of this cross section, in combination with experimental results, it may be possible to clearly identify the physical effects of saturation.
 
 One of the key innovations that enables this calculation is the hybrid formalism~\cite{Dumitru:2001jn}.
 In this approach, the gluon field of the target nucleus is represented by a correlator of Wilson lines.
 Unlike the traditional parton distributions of the parton model, Wilson lines can capture the physical effect of multiple gluon scattering by the projectile parton.
 The cross section then factorizes as
 \begin{equation}
  \frac{\dd[3]\sigma}{\dd\rapidity\dd[2]\vec{p}_\perp} = \sum_i\int\frac{\dd\zhadron}{\zhadron^2}\frac{\dd\xprojectile}{\xprojectile} \color{partondist}\xprojectile f_i(\xprojectile, \mu) \color{fragmentation}D_{h/i}(z, \mu) \color{gluondist}\dipoleF\biggl(\frac{p_\perp}{\zhadron}\biggr) \color{kinfactor}\mathcal{P}(\xi) (\ldots)
 \end{equation}
 where $f_i$ is a traditional parton distribution, $D_{h/i}$ is a fragmentation function, $\dipoleF$ is the momentum-space correlator of Wilson lines (also known as the unintegrated dipole gluon distribution), and $\mathcal{P}(\xi)$ and the remaining factors are the ``hard factors'' that describe the parton-level interaction, calculated from perturbative QCD.

 \section{History of the \pA{} Cross Section}
 
 \tikzset{NLO diagram at/.style 2 args={scale=0.5,xshift={5cm*#2},yshift={-3.6cm*#1}}}
 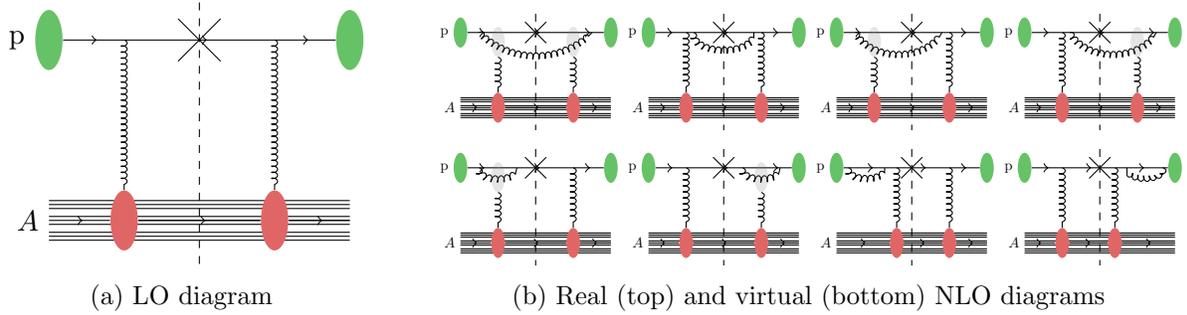
\begin{figure}
  \begin{subfigure}[b]{.3\textwidth}
   \centering
   \tikzsetnextfilename{diagrams-lo}
   \begin{tikzpicture}[
    every node/.append style=transform shape
   ]
    \begin{scope}[scale=1]
     \coordinate (origin) at (0,0);
     \node[blob] (left proton) at (-2,0) {};
     \node[blob] (left nucleus) at (-1,-2.4) {};
     \coordinate (left interaction) at (left nucleus |- origin);
     \coordinate (right interaction) at ($(left interaction)!2!(left interaction -| origin)$);
     \node[blob] (right nucleus) at ($(left nucleus)!2!(left nucleus -| origin)$) {};
     \node[blob] (right proton) at ($(left proton)!2!(left proton -| origin)$) {};

     \node[left=1ex] at (left proton) {$\Pproton$};
     \draw[quark] (left proton) to (left interaction);
     \draw[quark] (left interaction) to (right interaction);
     \draw[quark] (right interaction) to (right proton);
     \draw[nucleus] (left proton |- left nucleus) node[left] {$A$} to (left nucleus);
     \draw[nucleus] (left nucleus) to (right nucleus);
     \draw[nucleus] (right nucleus) to (right nucleus -| right proton);
     
     \draw[small gluon] (left interaction) to (left nucleus);
     \draw[small gluon] (right interaction) to (right nucleus);
     
     \node[fill=partondist!60!white,blob] at (left proton) {};
     \node[fill=gluondist!60!white,blob] at (left nucleus) {};
     \node[fill=partondist!60!white,blob] at (right proton) {};
     \node[fill=gluondist!60!white,blob] at (right nucleus) {};
     
     \draw[dashed] (origin) +(0,0.5) -- (origin |- left nucleus) -- +(0,-0.6);
     \draw (origin) +(45:.4) -- +(45:-.4) +(-45:.4) -- +(-45:-.4);
    \end{scope}
   \end{tikzpicture}
   \caption{LO diagram}
   \label{fig:diagrams:lo}
  \end{subfigure}
  \begin{subfigure}[b]{.7\textwidth}
   \centering
   \tikzsetnextfilename{diagrams-nlo}
   \begin{tikzpicture}[
    every node/.append style=transform shape
   ]
    \begin{scope}[NLO diagram at={1}{1}]
     \coordinate (origin) at (0,0);
     \node[blob] (left proton) at (-2,0) {};
     \node[blob] (left nucleus) at (-1,-2) {};
     \coordinate (left interaction) at (left nucleus |- origin);
     \coordinate (right interaction) at ($(left interaction)!2!(left interaction -| origin)$);
     \node[blob] (right nucleus) at ($(left nucleus)!2!(left nucleus -| origin)$) {};
     \node[blob] (right proton) at ($(left proton)!2!(left proton -| origin)$) {};

     \coordinate (left gluon attach) at ($(left proton)!.5!(left interaction)$);
     \coordinate (right gluon attach) at ($(right proton)!.5!(right interaction)$);
     
     \node[left=1ex] at (left proton) {$\Pproton$};
     \draw[quark] (left proton) to (left interaction);
     \draw[quark] (left interaction) to (right interaction);
     \draw[quark] (right interaction) to (right proton);
     \draw[nucleus] (left proton |- left nucleus) node[left] {$A$} to (left nucleus);
     \draw[nucleus] (left nucleus) to (right nucleus);
     \draw[nucleus] (right nucleus) to (right nucleus -| right proton);
     \draw[small gluon] (left gluon attach) to[out=-45,in=225] (right gluon attach);
     
     \coordinate (left gluon intermediate) at ($(left interaction)!.5!90:(left proton)$);
     \coordinate (right gluon intermediate) at ($(right interaction)!.5!270:(right proton)$){};
     \node[blob,fill=gray,fill opacity=0.2,fit=(left interaction) (left gluon intermediate)] (left gluon blob) {};
     \node[blob,fill=gray,fill opacity=0.2,fit=(right interaction) (right gluon intermediate)] (right gluon blob) {};
     \draw[small gluon] (left gluon blob) to (left nucleus);
     \draw[small gluon] (right gluon blob) to (right nucleus);
     
     \node[fill=partondist!60!white,blob] at (left proton) {};
     \node[fill=gluondist!60!white,blob] at (left nucleus) {};
     \node[fill=partondist!60!white,blob] at (right proton) {};
     \node[fill=gluondist!60!white,blob] at (right nucleus) {};
     
     \draw[dashed] (origin) +(0,0.5) -- (origin |- left nucleus) -- +(0,-0.6);
     \draw (origin) +(45:.4) -- +(45:-.4) +(-45:.4) -- +(-45:-.4);
    \end{scope}

    \begin{scope}[NLO diagram at={1}{2}]
     \coordinate (origin) at (0,0);
     \node[blob] (left proton) at (-2,0) {};
     \node[blob] (left nucleus) at (-1,-2) {};
     \coordinate (left interaction) at (left nucleus |- origin);
     \coordinate (right interaction) at ($(left interaction)!2!(left interaction -| origin)$);
     \node[blob] (right nucleus) at ($(left nucleus)!2!(left nucleus -| origin)$) {};
     \node[blob] (right proton) at ($(left proton)!2!(left proton -| origin)$) {};

     \coordinate (left gluon attach) at ($(left proton)!1.2!(left interaction)$);
     \coordinate (right gluon attach) at ($(right proton)!1.2!(right interaction)$);
     
     \node[left=1ex] at (left proton) {$\Pproton$};
     \draw[quark] (left proton) to (left interaction);
     \draw[quark] (left interaction) to (right interaction);
     \draw[quark] (right interaction) to (right proton);
     \draw[nucleus] (left proton |- left nucleus) node[left] {$A$} to (left nucleus);
     \draw[nucleus] (left nucleus) to (right nucleus);
     \draw[nucleus] (right nucleus) to (right nucleus -| right proton);
     \draw[small gluon] (left gluon attach) to[out=-90,in=270] (right gluon attach);
     
     \draw[small gluon] (left interaction) to (left nucleus);
     \draw[small gluon] (right interaction) to (right nucleus);
     
     \node[fill=partondist!60!white,blob] at (left proton) {};
     \node[fill=gluondist!60!white,blob] at (left nucleus) {};
     \node[fill=partondist!60!white,blob] at (right proton) {};
     \node[fill=gluondist!60!white,blob] at (right nucleus) {};

     \draw[dashed] (origin) +(0,0.5) -- (origin |- left nucleus) -- +(0,-0.6);
     \draw (origin) +(45:.4) -- +(45:-.4) +(-45:.4) -- +(-45:-.4);
    \end{scope}

    \begin{scope}[NLO diagram at={1}{3}]
     \coordinate (origin) at (0,0);
     \node[blob] (left proton) at (-2,0) {};
     \node[blob] (left nucleus) at (-1,-2) {};
     \coordinate (left interaction) at (left nucleus |- origin);
     \coordinate (right interaction) at ($(left interaction)!2!(left interaction -| origin)$);
     \node[blob] (right nucleus) at ($(left nucleus)!2!(left nucleus -| origin)$) {};
     \node[blob] (right proton) at ($(left proton)!2!(left proton -| origin)$) {};

     \coordinate (left gluon attach) at ($(left proton)!.5!(left interaction)$);
     \coordinate (right gluon attach) at ($(right proton)!1.2!(right interaction)$);
     
     \node[left=1ex] at (left proton) {$\Pproton$};
     \draw[quark] (left proton) to (left interaction);
     \draw[quark] (left interaction) to (right interaction);
     \draw[quark] (right interaction) to (right proton);
     \draw[nucleus] (left proton |- left nucleus) node[left] {$A$} to (left nucleus);
     \draw[nucleus] (left nucleus) to (right nucleus);
     \draw[nucleus] (right nucleus) to (right nucleus -| right proton);
     \draw[small gluon] (left gluon attach) to[out=-60,in=240] (right gluon attach);
     
     \coordinate (left gluon intermediate) at ($(left interaction)!.5!90:(left proton)$);
     \node[blob,fill=gray,fill opacity=0.2,fit=(left interaction) (left gluon intermediate)] (left gluon blob) {};
     \draw[small gluon] (left gluon blob) to (left nucleus);
     \draw[small gluon] (right interaction) to (right nucleus);
     
     \node[fill=partondist!60!white,blob] at (left proton) {};
     \node[fill=gluondist!60!white,blob] at (left nucleus) {};
     \node[fill=partondist!60!white,blob] at (right proton) {};
     \node[fill=gluondist!60!white,blob] at (right nucleus) {};

     \draw[dashed] (origin) +(0,0.5) -- (origin |- left nucleus) -- +(0,-0.6);
     \draw (origin) +(45:.4) -- +(45:-.4) +(-45:.4) -- +(-45:-.4);
    \end{scope}

    \begin{scope}[NLO diagram at={1}{4}]
     \coordinate (origin) at (0,0);
     \node[blob] (left proton) at (-2,0) {};
     \node[blob] (left nucleus) at (-1,-2) {};
     \coordinate (left interaction) at (left nucleus |- origin);
     \coordinate (right interaction) at ($(left interaction)!2!(left interaction -| origin)$);
     \node[blob] (right nucleus) at ($(left nucleus)!2!(left nucleus -| origin)$) {};
     \node[blob] (right proton) at ($(left proton)!2!(left proton -| origin)$) {};

     \coordinate (left gluon attach) at ($(left proton)!1.2!(left interaction)$);
     \coordinate (right gluon attach) at ($(right proton)!.5!(right interaction)$);
     
     \node[left=1ex] at (left proton) {$\Pproton$};
     \draw[quark] (left proton) to (left interaction);
     \draw[quark] (left interaction) to (right interaction);
     \draw[quark] (right interaction) to (right proton);
     \draw[nucleus] (left proton |- left nucleus) node[left] {$A$} to (left nucleus);
     \draw[nucleus] (left nucleus) to (right nucleus);
     \draw[nucleus] (right nucleus) to (right nucleus -| right proton);
     \draw[small gluon] (left gluon attach) to[out=-60,in=240] (right gluon attach);
     
     \draw[small gluon] (left interaction) to (left nucleus);
     \coordinate (right gluon intermediate) at ($(right interaction)!.5!270:(right proton)$);
     \node[blob,fill=gray,fill opacity=0.2,fit=(right interaction) (right gluon intermediate)] (right gluon blob) {};
     \draw[small gluon] (right gluon blob) to (right nucleus);
     
     \node[fill=partondist!60!white,blob] at (left proton) {};
     \node[fill=gluondist!60!white,blob] at (left nucleus) {};
     \node[fill=partondist!60!white,blob] at (right proton) {};
     \node[fill=gluondist!60!white,blob] at (right nucleus) {};

     \draw[dashed] (origin) +(0,0.5) -- (origin |- left nucleus) -- +(0,-0.6);
     \draw (origin) +(45:.4) -- +(45:-.4) +(-45:.4) -- +(-45:-.4);
    \end{scope}

    \begin{scope}[NLO diagram at={2}{1}]
     \coordinate (origin) at (0,0);
     \node[blob] (left proton) at (-2,0) {};
     \node[blob] (left nucleus) at (-1,-2) {};
     \coordinate (left interaction) at (left nucleus |- origin);
     \coordinate (right interaction) at ($(left interaction)!2!(left interaction -| origin)$);
     \node[blob] (right nucleus) at ($(left nucleus)!2!(left nucleus -| origin)$) {};
     \node[blob] (right proton) at ($(left proton)!2!(left proton -| origin)$) {};

     \coordinate (left gluon attach) at ($(left proton)!.5!(left interaction)$);
     \coordinate (right gluon attach) at ($(left gluon attach)!2!(left interaction)$);
     
     \node[left=1ex] at (left proton) {$\Pproton$};
     \draw[quark] (left proton) to (left interaction);
     \draw[quark] (left interaction) to (right interaction);
     \draw[quark] (right interaction) to (right proton);
     \draw[nucleus] (left proton |- left nucleus) node[left] {$A$} to (left nucleus);
     \draw[nucleus] (left nucleus) to (right nucleus);
     \draw[nucleus] (right nucleus) to (right nucleus -| right proton);
     \draw[small gluon] (left gluon attach) to[out=-90,in=270] (right gluon attach);
     
     \coordinate (left gluon intermediate) at ($(left interaction)!.5!90:(left proton)$);
     \node[blob,fill=gray,fill opacity=0.2,fit=(left interaction) (left gluon intermediate)] (left gluon blob) {};
     \draw[small gluon] (left gluon blob) to (left nucleus);
     \draw[small gluon] (right interaction) to (right nucleus);
     
     \node[fill=partondist!60!white,blob] at (left proton) {};
     \node[fill=gluondist!60!white,blob] at (left nucleus) {};
     \node[fill=partondist!60!white,blob] at (right proton) {};
     \node[fill=gluondist!60!white,blob] at (right nucleus) {};
     
     \draw[dashed] (origin) +(0,0.5) -- (origin |- left nucleus) -- +(0,-0.6);
     \draw (origin) +(45:.4) -- +(45:-.4) +(-45:.4) -- +(-45:-.4);
    \end{scope}

    \begin{scope}[NLO diagram at={2}{2}]
     \coordinate (origin) at (0,0);
     \node[blob] (left proton) at (-2,0) {};
     \node[blob] (left nucleus) at (-1,-2) {};
     \coordinate (left interaction) at (left nucleus |- origin);
     \coordinate (right interaction) at ($(left interaction)!2!(left interaction -| origin)$);
     \node[blob] (right nucleus) at ($(left nucleus)!2!(left nucleus -| origin)$) {};
     \node[blob] (right proton) at ($(left proton)!2!(left proton -| origin)$) {};

     \coordinate (right gluon attach) at ($(right proton)!.5!(right interaction)$);
     \coordinate (left gluon attach) at ($(right gluon attach)!2!(right interaction)$);
     
     \node[left=1ex] at (left proton) {$\Pproton$};
     \draw[quark] (left proton) to (left interaction);
     \draw[quark] (left interaction) to (right interaction);
     \draw[quark] (right interaction) to (right proton);
     \draw[nucleus] (left proton |- left nucleus) node[left] {$A$} to (left nucleus);
     \draw[nucleus] (left nucleus) to (right nucleus);
     \draw[nucleus] (right nucleus) to (right nucleus -| right proton);
     \draw[small gluon] (left gluon attach) to[out=-90,in=270] (right gluon attach);
     
     \coordinate (right gluon intermediate) at ($(right interaction)!.5!270:(right proton)$);
     \node[blob,fill=gray,fill opacity=0.2,fit=(right interaction) (right gluon intermediate)] (right gluon blob) {};
     \draw[small gluon] (right gluon blob) to (right nucleus);
     \draw[small gluon] (left interaction) to (left nucleus);
     
     \node[fill=partondist!60!white,blob] at (left proton) {};
     \node[fill=gluondist!60!white,blob] at (left nucleus) {};
     \node[fill=partondist!60!white,blob] at (right proton) {};
     \node[fill=gluondist!60!white,blob] at (right nucleus) {};
     
     \draw[dashed] (origin) +(0,0.5) -- (origin |- left nucleus) -- +(0,-0.6);
     \draw (origin) +(45:.4) -- +(45:-.4) +(-45:.4) -- +(-45:-.4);
    \end{scope}

    \begin{scope}[NLO diagram at={2}{3}]
     \coordinate (origin) at (0,0);
     \node[blob] (left proton) at (-2,0) {};
     \node[blob] (left nucleus) at (-0.4,-2) {};
     \coordinate (left interaction) at (left nucleus |- origin);
     \node[blob] (right nucleus) at (1,-2) {};
     \coordinate (right interaction) at (right nucleus |- origin);
     \node[blob] (right proton) at ($(left proton)!2!(left proton -| origin)$) {};

     \coordinate (left gluon attach) at ($(left proton)!.2!(left interaction)$);
     \coordinate (right gluon attach) at ($(left proton)!.8!(left interaction)$);
     
     \node[left=1ex] at (left proton) {$\Pproton$};
     \draw[quark] (left proton) to (left interaction);
     \draw[quark] (left interaction) to (right interaction);
     \draw[quark] (right interaction) to (right proton);
     \draw[nucleus] (left proton |- left nucleus) node[left] {$A$} to (left nucleus);
     \draw[nucleus] (left nucleus) to (right nucleus);
     \draw[nucleus] (right nucleus) to (right nucleus -| right proton);
     \draw[small gluon] (left gluon attach) to[out=-90,in=270] (right gluon attach);
     
     \draw[small gluon] (left interaction) to (left nucleus);
     \draw[small gluon] (right interaction) to (right nucleus);
     
     \node[fill=partondist!60!white,blob] at (left proton) {};
     \node[fill=gluondist!60!white,blob] at (left nucleus) {};
     \node[fill=partondist!60!white,blob] at (right proton) {};
     \node[fill=gluondist!60!white,blob] at (right nucleus) {};
     
     \draw[dashed] (origin) +(0,0.5) -- (origin |- left nucleus) -- +(0,-0.6);
     \draw (origin) +(45:.4) -- +(45:-.4) +(-45:.4) -- +(-45:-.4);
    \end{scope}

    \begin{scope}[NLO diagram at={2}{4}]
     \coordinate (origin) at (0,0);
     \node[blob] (left proton) at (-2,0) {};
     \node[blob] (left nucleus) at (-1,-2) {};
     \coordinate (left interaction) at (left nucleus |- origin);
     \node[blob] (right nucleus) at (0.4,-2) {};
     \coordinate (right interaction) at (right nucleus |- origin);
     \node[blob] (right proton) at ($(left proton)!2!(left proton -| origin)$) {};

     \coordinate (left gluon attach) at ($(right proton)!.2!(right interaction)$);
     \coordinate (right gluon attach) at ($(right proton)!.8!(right interaction)$);
     
     \node[left=1ex] at (left proton) {$\Pproton$};
     \draw[quark] (left proton) to (left interaction);
     \draw[quark] (left interaction) to (right interaction);
     \draw[quark] (right interaction) to (right proton);
     \draw[nucleus] (left proton |- left nucleus) node[left] {$A$} to (left nucleus);
     \draw[nucleus] (left nucleus) to (right nucleus);
     \draw[nucleus] (right nucleus) to (right nucleus -| right proton);
     \draw[small gluon] (left gluon attach) to[out=-90,in=270] (right gluon attach);
     
     \draw[small gluon] (left interaction) to (left nucleus);
     \draw[small gluon] (right interaction) to (right nucleus);
     
     \node[fill=partondist!60!white,blob] at (left proton) {};
     \node[fill=gluondist!60!white,blob] at (left nucleus) {};
     \node[fill=partondist!60!white,blob] at (right proton) {};
     \node[fill=gluondist!60!white,blob] at (right nucleus) {};
     
     \draw[dashed] (origin) +(0,0.5) -- (origin |- left nucleus) -- +(0,-0.6);
     \draw (origin) +(45:.4) -- +(45:-.4) +(-45:.4) -- +(-45:-.4);
    \end{scope}

   \end{tikzpicture}
   \caption{Real (top) and virtual (bottom) NLO diagrams}
   \label{fig:diagrams:nlo}
  \end{subfigure}
  \caption{Diagrams contributing to the quark-quark channel --- with a quark projectile, and a quark fragmenting into the final-state hadron --- at LO and NLO. This is one of four channels.}
  \label{fig:diagrams}
 \end{figure}

 The first calculation of the inclusive hadron cross section in proton-nucleus collisions was done by Dumitru and Jalilian-Marian in 2002~\cite{Dumitru:2002qt}.
 Their result describes the process shown in figure~\ref{fig:diagrams:lo} in which a quark from the projectile proton interacts with the gluon field of the nucleus, as well as the equivalent process with a gluon projectile instead of a quark.
 Shortly afterwards, they supplemented this with a numerical calculation~\cite{Dumitru:2005gt} showing that the hybrid model is able to describe experimental results from RHIC, up to a $\pperp$-independent factor $K$.
 The $K$ factor was expected to approximately account for the missing next-to-leading order contributions.
 
 Extending the calculation to next-to-leading order began with the inelastic terms, roughly corresponding to real diagrams in which the emitted gluon persists into the final state, as shown in the top row of figure~\ref{fig:diagrams:nlo}.
 Albacete \etal. computed the corresponding expressions and obtained numerical results in reference~\cite{Albacete:2012xq} with two candidate gluon distributions.
 They were able to match data from the RHIC experiments at low $\pperp$ with $K=1$ for charged hadrons~\cite{Arsene:2004ux} and $K=0.4$ for neutral hadrons~\cite{Adams:2006uz}.
 However, the calculated result drops sharply at high $\pperp$.
 
\begin{figure}
  \begin{subfigure}[b]{.55\textwidth}
   \tikzsetnextfilename{brahms-hiY}
   \begin{tikzpicture}
    \begin{axis}[
     xtick placement tolerance=-0.05pt,
     result axis,
     width=7cm,
     height=6cm,
     ymode=log,
     ymin=8e-8,
     ymax=5e1,
     xmin=0,
     xmax=3.5,
     title style={align=center},
    ]
     \resultplot{rcBK LO,brahms xsec LO}{brahmsrcBKB}
     \resultplot{rcBK NLO,brahms xsec NLO}{brahmsrcBKB}
     \resultplot{rcBK extra,brahms xsec Lqg}{brahmsrcBKB}
     \addplot[data plot,brahms data] table {\brahmsdAuhiY};
     \legend{$\text{LO}$,$+\text{NLO}$,$+L_q+L_g$,BRAHMS}
     \node[anchor=south west] at (axis description cs:0.03,0.03) {$y = 3.2$};
    \end{axis}
   \end{tikzpicture}
   \caption{BRAHMS, $\mu^2 = \text{\SIrange{10}{50}{GeV^2}}$}
   \label{fig:results:brahms}
  \end{subfigure}
  \begin{subfigure}[b]{.45\textwidth}
   \tikzsetnextfilename{atlas-hiY}
   \begin{tikzpicture}
    \begin{axis}[
      xtick placement tolerance=-0.05pt,
      result axis,
      width=7cm,
      height=6cm,
      ymode=log,
      ymin=1e-6,
      ymax=4e1,
      xmin=0,
      xmax=7,
      ylabel={},
      title style={align=center},
    ]
     \resultplot{rcBK LO,atlas xsec LO}{atlasrcBKB}
     \resultplot{rcBK NLO,atlas xsec NLO}{atlasrcBKB}
     \resultplot{rcBK extra,atlas xsec Lqg}{atlasrcBKB}
     \addplot[data plot,atlas y175 data] table {\atlaspPb};
     \legend{$\text{LO}$,$+\text{NLO}$,$+L_q+L_g$,ATLAS}
     \node[anchor=south west] at (axis description cs:0.03,0.03) {$\rapidity = 1.75$};
    \end{axis}
   \end{tikzpicture}
   \caption{ATLAS, $\mu^2 = \text{\SIrange{10}{100}{GeV^2}}$}
   \label{fig:results:atlas}
  \end{subfigure}
  \caption{Calculated results~\cite{Stasto:2013cha,Watanabe:2015tja} for forward rapidity at BRAHMS and ATLAS, showing LO as well as NLO without and with the kinematical constraint ($L_q + L_g$). The error band shows the results for two values of $\mu^2$, as indicated. Calculations were done with the running coupling BK gluon distribution; data are from BRAHMS~\cite{Arsene:2004ux} and ATLAS~\cite{AlexanderMilovonbehalfoftheATLAS:2014rta} are also shown for comparison. Figures from reference~\cite{Watanabe:2015tja}.}
  \label{fig:results}
 \end{figure}
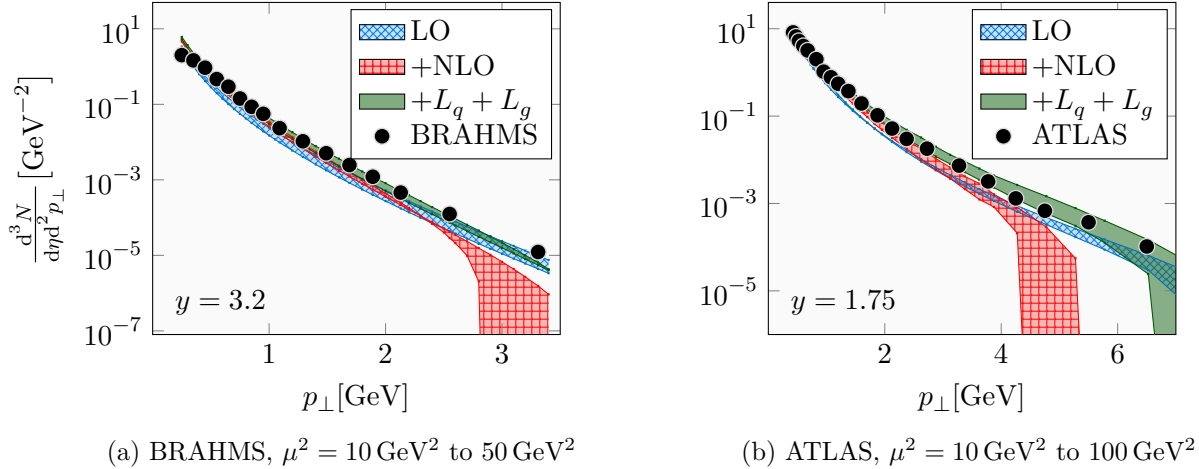

 It was natural to wonder whether the inclusion of the virtual diagrams, such as those shown in figure~\ref{fig:diagrams:nlo}, would bolster the results at high $\pperp$.
 The theoretical calculation of the corresponding amplitudes was performed by Chirilli, Xiao, and Yuan~\cite{Chirilli:2012jd}, and numerical results were first obtained shortly thereafter by Sta\'sto, Xiao, and Zaslavsky~\cite{Stasto:2013cha}.
 As shown by the sample results in figure~\ref{fig:results}, including the full NLO corrections does improve the agreement with experimental data at low $\pperp$.
 It also renders the $K$ factor unnecessary for both charged and neutral hadrons.
 But the most evident feature is that the drop at high $\pperp$ still exists after incorporating the virtual diagrams; in fact, the cross section becomes negative!

 This conclusion touched off a flurry of speculation about the significance of a negative predicted cross section, and how (or if) it might be possible to modify the calculation to produce a uniformly positive prediction~\cite{Kang:2014lha,Xiao:2014uba,Stasto:2014sea,Altinoluk:2014eka,Watanabe:2015tja}.
 However, calculations with a variety of gluon distributions and kinematic conditions~\cite{Stasto:2013cha,Zaslavsky:2014asa} show that the negativity appears to be a fairly fundamental feature of the NLO result.
 In particular, it seems not to be a consequence of the choice of the candidate gluon distribution.

 Most recently, references~\cite{Altinoluk:2014eka,Watanabe:2015tja} have examined the effect of a kinematical constraint on the emitted gluon momentum fraction $\xi$.
 This constraint, which limits the maximum value of $\xi$, can be incorporated into the calculation by modification of the dipole splitting function, and it results in additional terms, designated $L_q$ and $L_g$ in figure~\ref{fig:results}, that contribute at the NLO level.
 As shown in the figures, the new terms extend the $\pperp$ range over which the calculation agrees with the experimental results, but they do not cure the negativity.

 The results of the calculation for ATLAS parameters, computed for the first time in reference~\cite{Watanabe:2015tja}, are particularly interesting because the LHC is able to reach smaller values of $\xtarget$ than RHIC, and is better positioned to probe the kinematic regime where saturation becomes relevant.
 Obtaining these numerical results required overcoming a number of technical challenges.
 The greatly expanded phase space induces large statistical fluctuations in Monte Carlo integration, and it becomes harder to recover the precise cancellations between $\order{\kperp^{-2}}$ terms that are needed for the results to work out properly.
 In addition, the larger momenta accessible to an LHC collision mean that numerical Fourier transforms are no longer viable, as they were at RHIC.
 It's necessary to convert all expressions into momentum space, which required new Fourier transform formulas.
 Now that the formulas have been adapted to LHC conditions, it opens up a wide variety of hybrid formalism calculations that can be computed and compared to LHC data to come.
 
 \section{Conclusion}
 With the (hopefully) complete next-to-leading order calculation of the $\HepProcess{\Pproton\Pnucleus\to\Phadron + \Panything}$ cross section~\cite{Chirilli:2012jd,Watanabe:2015tja} at hand, and a stable numerical implementation~\cite{Stasto:2013cha,Zaslavsky:2014asa} available, we can look forward to precise numerical predictions for both RHIC and LHC measurements of the cross section.
 These predictions will be critical for taking advantage of the LHC's unprecedented kinematic reach to probe directly into the saturation regime, and perhaps detect the direct effects of saturation and the color glass condensate.

 However, in order to have a truly believable signal of saturation, it will be necessary to test a range of predictions.
 We hope to see additional data on the inclusive hadron yield, especially at higher rapidities, from the LHC in the near future.
 If the hybrid model with a suitable candidate gluon distribution is able to predict the variation in inclusive hadron yield with rapidity, it will be a very strong indicator that the model is on the right track.
 
 \printbibliography
\end{document}